\documentclass[prd,aps,twocolumn,showpacs]{revtex4}

\usepackage[dvips]{graphicx}
\usepackage{amsmath}

\begin{document}

\title{Continuum versus periodic lattice Monte Carlo approach
                to classical field theory}

\author{Bogdan Mihaila}
\email{bogdan.mihaila@unh.edu} \affiliation{Physics Division,
   Argonne National Laboratory, Argonne, IL 60439}

\author{John~F.~Dawson}
\email{john.dawson@unh.edu} \affiliation{Department of Physics,
   University of New Hampshire, Durham, NH 03824}
\date{\today}
\begin{abstract}
We compare the momentum space with the standard periodic lattice
approach to Monte Carlo calculations in classical $\phi^4$ field
theory. We show that the mismatch in the initial value of
$\phi^2_{\text{cl}}(t)$, results in a shift in the ``thermalized''
value, at large times. The two approaches converge to the same
result in the continuum limit.
\end{abstract}
\maketitle

\section{Introduction}

Classical field theory in 1+1 dimensions has been used recently as
a test ground for various approximation methods for calculating
the dynamical evolution of a nonequilibrium
system~\cite{ref:abw,ref:bva_cl,{ref:newberges}}. In classical
field theory one can numerically simulate the full time evolution
of the system via Monte Carlo methods~\cite{ref:abw}. Given the
probability distributions in the initial state of the system, one
can randomly sample the initial distributions and numerically
solve the equations of motion. The time evolution of the various
expectation values is then obtained by averaging over the number
of independent realizations. The dynamics becomes exact in the
limit when the number of initial samples goes to infinity.

It is important to be able to carry out calculations both in
coordinate (lattice) and momentum space. In a lattice calculation,
operators are numerically realized by using finite-difference
representations of derivatives and integrals. Finite-difference
methods, though leading to sparse matrices, are notoriously slow
to converge. Thus the need of using higher order methods, like the
nonuniform grid Chebyshev polynomial
methods~\cite{ref:BMIM,ref:MDC} we used in the past few years,
which belong to the class of spectral methods. Then the resulting
matrices are less sparse, but what we apparently loose in storage
requirements, we gain in speed. We do in fact keep the storage
needs moderate also, as we can achieve very good accuracy with a
moderate number of grid sites. Spectral methods have made possible
for the first time to carry out complex dynamical calculations at
next to leading order in quantum
mechanics~\cite{ref:MDC,ref:paper1,ref:us1} and field
theory~\cite{ref:bva_cl}.

In order to evaluate the quality of the various approximations,
one needs to find ways of comparing the truncated dynamics
predicted by a given approximation scheme with the exact result.
Hence the importance of classical field theory where Monte Carlo
methods allow us to probe the true dynamics of a hamiltonian
system. In practice, the exact calculation is always carried out
in two steps: First, we consider a truncated space, which is
characterized by a cutoff in momentum space. Once a solution is
obtained for a given finite value of the cutoff, we increase the
value of the cutoff and repeat the calculation until the result
becomes independent of the cutoff. This second step is what we
usually refer to as \emph{taking the continuum limit}.

Lattice and momentum space (continuum) calculations do give the
same result in the continuum limit. However, in a truncated space
the results will be different. Since we are trying to compare the
exact dynamics of a system with the result of a non-lattice based
calculation, we cannot reliably use the lattice result. Hence the
need for a continuum implementation of the Monte Carlo method.

It is interesting to remark here that the study of the classical
$\phi^4$ field theory problem represents a warmup for the
next-to-leading order studies of the quantum linear sigma
model~\cite{ref:lin_sig} and beyond. For the linear sigma model,
the presence of the Landau pole will prevent us from taking the
momentum cutoff to arbitrarily large values. Therefore, we are in
fact less interested in the continuum limit of the problem, at
least for now.

%
%
\section{The Lagrangian}

For a real field $\phi(x)$, the classical Lagrangian density in
the $\phi^4(x)$ theory is given by:
\begin{align*}
   & \mathcal{L}(\phi,\partial_\mu \phi)
    =
   \frac{1}{2} \, [ \partial_\mu \phi(x) ][ \partial^\mu \phi(x) ]
   -
   \frac{\lambda}{8}
   \left  [
      \phi^2(x) - r_0^2
   \right ]^{2}  \>,
\end{align*}
or
\begin{align}
   & \mathcal{L}(\phi,\partial_\mu \phi)
   \label{e:Lagi}
   \\ \notag
   & =
   \frac{1}{2} \,
      \left  \{
         ( \partial_\mu \phi(x) ) ( \partial^\mu \phi(x) )
         -
         \mu^2 \, \phi^2(x)
      \right \}
   -
   \frac{\lambda}{8} \, \phi^4(x)
   -
   \frac{\mu^4 }{2 \lambda} \>,
\end{align}
where $\mu^2 = - \lambda r_0^2 / 2 > 0$. The classical equation of
motion for $\phi(x)$, in 1+1 dimensions, is:
\begin{equation}
   \left  [ \,
      \partial_{t}^2
      -
      \partial_{x}^2
      +
      \mu^2 + \frac{\lambda}{2} \, \phi^2(x,t)
   \right ] \, \phi(x,t)
   = 0 \>.
\label{e:motion}
\end{equation}
In order to make contact with the work of Aarts \emph{et
al}~\cite{ref:abw}, in this paper we address only the symmetric
case, $\langle \phi(0) \rangle = \langle \pi(0) \rangle = 0$.

%
%
\section{Initial values}

Assuming that initially the system is in thermal equilibrium, the
initial values $\phi(x,0)$ and $\pi(x,0) = \dot \phi(x,0)$ are
taken from a canonical ensemble governed by a classical density
distribution $\rho[\phi,\pi]$ defined as
\begin{align}
   \rho[\phi,\pi]
   &=
   Z^{-1}(\beta_0) \, e^{ -\beta_0 H[\phi,\pi] } \>,
   \\ \notag
   Z(\beta_0)
   &=
   \prod_x \int \int
   {\mathrm  d} \phi(x) \, {\mathrm  d} \pi(x) \,
   e^{ -\beta_0 H[\phi,\pi] }  \>.
\end{align}
with $\beta_0 = 1/T_0$, and $\pi(x,t) = \dot \phi(x,t)$.
Correspondingly, the ensemble average of a quantity $A[\phi,\pi]$
is defined by
\begin{align}
   \langle A[\phi,\pi] \rangle
   & =
   \mathrm{Tr} \{ \rho A[\phi,\pi] \}
   \label{e:expectfull}
\end{align}

Following Aarts \emph{et al}~\cite{ref:abw}, we choose initial
values by randomly sampling the density distribution corresponding
to the free particle Hamiltonian ($\lambda=0$). For each set of
initial conditions, we time-evolve $\phi(x,t)$ using the equation
of motion~(\ref{e:motion}). The average value
$\phi^2_{\text{cl}}(t)$ is calculated as
\begin{equation}
   \phi^2_{\text{cl}}(t)
   =
   \frac{1}{M_c} \ \sum_{i=1}^{M_c} \
   \lim_{L \rightarrow \infty}
   \frac{1}{L} \ \int_{-\frac{L}{2}}^{\frac{L}{2}} \
   \phi^2(x,t) \ {\mathrm d} x
   \>,
   \label{e:phi2clave}
\end{equation}
where $(i)$ denotes the $i^{\text{th}}$ Monte-Carlo run and $M_c$
is the total number of Monte-Carlo runs.

%
%
\section{Momentum Space approach}

In the momentum space approach one introduces the Fourier
transform of the field $\tilde \phi(q,t)$ via
\begin{equation*}
   \phi(x,t)
   =
   \int \frac{ {\mathrm  d}q }{2\pi} \
   \tilde \phi(q,t) \, e^{- {\mathrm i} q x } \>.
\end{equation*}
One then obtains the classical equation of motion in momentum
space
\begin{equation*}
   \Bigl [ \,
      \partial_{t}^2
      \ + \left ( q^2 + \mu^2 \right )
   \Bigr ] \, \tilde \phi(q,t)
      + \frac{\lambda}{2} \,
        \int {\mathrm  d}x \ e^{{\mathrm i} q x} \ \phi^3(x,t)
   = 0 \>.
\end{equation*}
The equation of motion is solved by using an Euler method, where
the time-differential operator is replaced by a second order
difference formula, i.e.
\begin{equation*}
   \partial^2_t \tilde \phi(q,t)
   \rightarrow
   \Bigl [ \tilde \phi(q,t+\tau)
           - 2 \, \tilde \phi(q,t) + \tilde \phi(q,t-\tau)
   \Bigr ] \, / \, \tau^2
   \>,
\end{equation*}
where $\tau$ is the time step. A momentum space cut-off is used
and the system is also in a box of length $L$.  Letting $L= Na$,
and choosing our momentum to be
\begin{equation}
q=\frac{2\pi k}{L}, \;\;\;\;k=\left\{-\frac{N}{2}, \ldots,
\frac{N}{2}-1 \right\} \>, \label{eq:qk}
\end{equation}
we then have that
\begin{equation}
   \Lambda = \frac{\pi N}{L} = \frac{\pi}{a}
   \>,
\label{eq:limits}
\end{equation}
where  $\Lambda$ is the
momentum cutoff and $a$ is the lattice spacing.  Then, we must
take the continuum limit as
\begin{equation*}
   \int \frac{ {\mathrm  d}q }{2\pi} \bigl ( \ \bigr )
   \rightarrow
   \lim_{\Lambda \rightarrow \infty}
       \int_{-\Lambda}^{\Lambda}
       \frac{ {\mathrm  d}q }{2\pi} \bigl ( \ \bigr )
   \>.
\end{equation*}
In this approach, the periodic boundary assumption is implied
whenever we perform numerical Fourier transforms and convolutions,
the continuum Fourier being replaced by the discrete one.

We will choose the initial values for $\tilde \phi(q,0)$ and
$\tilde \pi(q,0)$ to be solutions of the unperturbed Hamiltonian
\begin{equation}
   H_0
   =
   \frac{1}{2}
   \int {\mathrm d} x \,
   \biggl \{
      \pi_0^2(x) + \bigl [ \partial_x \phi_0(x) \bigr ]^2
      +
      \mu^2 \, \phi_0^2(x)
   \biggr \} \>,
   \label{e:H0classical}
\end{equation}
where $\pi_0(x,t) = \dot\phi_0(x,t)$.  The corresponding equation
of motion in momentum space is
\begin{equation}
   \left  [ \,
      \partial_{t}^2
      +
      \omega_q^2
   \right ] \, \tilde \phi_0(q,t)
   = 0 \>,
\label{eq:freede}
\end{equation}
with the dispersion relation
\begin{equation}
   \omega_q^2 \ = \ q^2 + \mu^2  \>.
   \label{e:omegak}
\end{equation}
General solutions of Eq.~\eqref{eq:freede} are of the form:
\begin{eqnarray*}
   \tilde \phi_0(q,t)
   & = &
   \frac{1}{\sqrt{2 \omega_q}} \
   \Bigl [
      \tilde{a}_q \, e^{-{\mathrm i}\omega_q t} \ + \
      \tilde{a}^\ast_{-q} \, e^{{\mathrm i}\omega_q t}
   \Bigr ]  \>,
\end{eqnarray*}
and
\begin{eqnarray*}
   \tilde \pi_0(q,t)
   =
   \dot{\tilde \phi}_0(q,t)
   & = &
   \frac{1}{i} \sqrt{ \frac{\omega_q}{2} } \
   \Bigl [
      \tilde{a}_q \, e^{-{\mathrm i}\omega_q t} \ - \
      \tilde{a}^\ast_{-q} \, e^{{\mathrm i}\omega_q t}
   \Bigr ]  \>.
\end{eqnarray*}
We can now calculate the classical density matrix as
\begin{align}
   \rho_0[a_m,a_m^{\ast}]
   &=
   \frac{1}{Z_0(\beta_0)} \,
   \prod_k e^{-\beta_0 \omega_{q_k} \, ( x_{q_k}^2 + y_{q_k}^2 ) / L }
   \>,
   \label{e:densityzi}
   \\ \notag
   Z_0(\beta_0)
   &=
   \prod_x \int \int
   {\mathrm  d} \phi_0(x) \, {\mathrm  d} \pi_0(x) \,
   e^{ -\beta_0 H_0[\phi_0,\pi_0] }  \>.
\end{align}
where we have put $a_{q_k} = x_{q_k} + {\mathrm i} y_{q_k}$.
Hence, the symmetric case scenario, $\langle \phi(0) \rangle =
\langle \pi(0) \rangle = 0$, implies that $x_{q_k}$ and $y_{q_k}$
are uniform random deviates between $0.0$ and $1.0\ $.

Using Eq.~(\ref{e:expectfull}) with the density distribution given
by (\ref{e:densityzi}), we obtain
\begin{eqnarray*}
   \langle \tilde{a}_{q_k} \tilde{a}_{q_{k'}} \rangle
   & = &
    \langle \tilde{a}_{q_k}^{\ast} \tilde{a}_{q_{k'}}^{\ast} \rangle
   \ = \
   0  \>,
   \\
   \langle \tilde{a}_{q_k}^{\ast}
           \tilde{a}_{q_{k'}}^{\phantom\ast} \rangle
   & = &
   \langle \tilde{a}_{q_k}
           \tilde{a}_{q_{k'}}^{\phantom\ast} \rangle
   \ = \
   n_{q_k}(\beta_0) \
   \delta_{k,k'} / L
   \>,
\end{eqnarray*}
where $n_q(\beta_0) = 1/(\beta_0 \omega_q)$.  Note that
$n_q(\beta_0)$ is the high temperature limit of the classical
Bose-Einstein occupation number distribution. Finally, we obtain
\begin{eqnarray*}
   \langle \phi^2(0) \rangle
   & = &
   \frac{1}{\beta_0} \
   \frac{1}{L} \sum_k \ \frac{1}{\omega_{q_k}^2}
   \ = \
   \frac{1}{\mu \beta_0} \
   I(\mu,\Lambda) \>,
   \\
   \langle \pi^2(0) \rangle
   & = &
   \frac{1}{\beta_0} \
   \frac{1}{L} \sum_k
   \ = \
   \frac{1}{\beta_0} \
   \frac{1}{a}
   \>,
\end{eqnarray*}
with
\begin{eqnarray}
   I(\mu,\Lambda)
   \ = \ \frac{1}{\pi} \ \arctan(\Lambda/\mu)
   \>.
\end{eqnarray}
Thus, for a given cut-off in momentum space, we have
\begin{eqnarray}
   \langle \phi^2(0) \rangle
   & = &
   \frac{1}{\mu \beta_0 \pi} \
   \arctan(\Lambda/\mu)
   \>.
   \label{eq:phi2_ini}
\end{eqnarray}
In the limit when $\Lambda \rightarrow \infty$, we obtain $\langle
\phi^2(0) \rangle = 1 / (2 \mu \beta_0)$.

%
%
\section{Periodic lattice approach}

For the lattice formalism, we follow closely the approach
presented in Reference~\cite{ref:abw}. We discretize the continuum
equations by using a lattice in coordinate space with spacing $a$
and periodic boundary conditions. The differential operators are
replaced by second order difference formulas:
\begin{align*}
   \partial^2_x \phi(x,t)
   &\rightarrow
   \Bigl [ \phi(x+a,t)
           - 2 \, \phi(x,t) + \phi(x-a,t)
   \Bigr ] \, / \, a^2
   \>,
   \\
   \partial^2_t \phi(x,t)
   &\rightarrow
   \Bigl [ \phi(x,t+\tau)
           - 2 \, \phi(x,t) + \phi(x,t-\tau)
   \Bigr ] \, / \, \tau^2
   \>.
\end{align*}
The dispersion relation is modified due to the Laplacian on the
lattice and becomes
\begin{equation}
\omega^2_{\hat q} = \hat q^2+\mu^2, \;\;\;\; \hat q^2 =
\frac{2}{a^2} (1-\cos aq)\>,
   \label{e:omegak_lattice}
\end{equation}
with $a=\pi / \Lambda$. The momentum $q$ takes the same finite
number of discrete values, see Eq.~\eqref{eq:qk}. The relationship
between the cutoffs in the two approaches is given by
\begin{equation*}
   \lim_{L \rightarrow \infty}
       \int_{-\frac{L}{2}}^{\frac{L}{2}} \bigl ( \ \bigr )
   \rightarrow
   \lim_{N \rightarrow \infty}
       \frac{1}{N a} \sum_{k=-\frac{N}{2}}^{\frac{N}{2}-1} \
       \bigl ( \ \bigr )
   \rightarrow
   \lim_{\Lambda \rightarrow \infty}
       \int_{-\Lambda}^{\Lambda}
       \frac{ {\mathrm  d}q }{2\pi} \bigl ( \ \bigr )
   \>.
\end{equation*}

The initial conditions are generated by sampling the initial
probability distribution of the unperturbed system just as in the
continuum case, with the formal difference that we replace the
values of the momenta $q_k$ by the shifted values~$\hat q_k$.
Then, the initial expectation value becomes
\begin{eqnarray*}
   \langle \phi^2(0) \rangle
   & = &
   \frac{1}{\beta_0} \
   \frac{1}{L} \sum_k \ \frac{1}{\omega_{\hat q_k}^2}
   \ = \
   \frac{1}{\mu \beta_0} \
   \hat I(\mu,a) \>,
\end{eqnarray*}
where
\begin{eqnarray}
   \hat I(\mu,a)
   \ = \ \frac{1}{2 \sqrt{1 + (\mu a/2)^2} }
   \>.
\end{eqnarray}
Thus, for a given lattice spacing $a=\pi/\Lambda$, we obtain
\begin{eqnarray}
   \langle \phi^2(0) \rangle
   & = &
   \frac{1}{\mu \beta_0} \
   \frac{1}{2 \sqrt{1 + (\mu a/2)^2} }
   \>.
   \label{eq:phi2_ini_lat}
\end{eqnarray}
In the limit when $a \rightarrow 0\  (\Lambda \rightarrow
\infty)$, we recover the continuum limit $\langle \phi^2(0)
\rangle = 1 / (2 \mu \beta_0)$.
\begin{figure}[!h]
   \centering
   \includegraphics[width=3.0in]{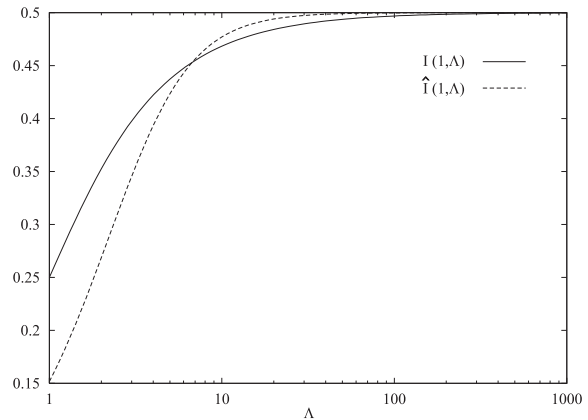}
   \caption{$I(\mu=1,\Lambda)$ and $\hat I(\mu=1,\Lambda)$ as a
   function of $\Lambda$.}
   \label{f:iml}
\end{figure}
However, for a given momentum space cutoff $\Lambda$, the values
of $\langle \phi^2(0) \rangle$ given by Eqs.~\eqref{eq:phi2_ini}
and~\eqref{eq:phi2_ini_lat} are not the same (see
Fig.~\ref{f:iml}). This is an artifact of solving the equations of
motions on the lattice and requiring $\phi(x,t)$ to satisfy
periodic boundary conditions. As a consequence, one cannot
directly compare continuum and lattice calculations.

%
%
\section{Results}

We choose to illustrate the approaches presented above, for a set
of parameters which allows us to compare with results available in
the literature~\cite{ref:abw}. We have $\lambda = 1/3$, $\mu=1$,
$\Lambda = 4 \pi$, $T_0=5.03891094$. Then, the initial condition
corresponding to the momentum space approach is $\langle \phi^2(0)
\rangle = 2.39208677$, as obtained from Eq.~\eqref{eq:phi2_ini},
while the numerically calculated initial average is
$\phi^2_{\text{cl}}(0) = 2.39187089$, for a 0.009\% error. In turn
the lattice calculation produces the continuum limit $\langle
\phi^2(0) \rangle = 2.5$, provided that one choose the lattice
spacing $a=0.25$. The lattice spacing is subsequently left
unchanged, even though one may vary the number of lattice sites,
and implicitly the lattice size (see Eq.~{\eqref{eq:limits}).

The numerical methods used for the numerical implementation of the
two methods are well under control. In Fig.~\ref{f:phi} we show
the Monte-Carlo value of $\phi^2_{\text{cl}}(t)$, as defined in
Eq.~\eqref{e:phi2clave}, obtained using the periodic lattice
approach. The error lines represent coordinate average deviations
of the runs as a function of $t$. Similar results are obtained
using the momentum space approach.
\begin{figure}
   \centering
   \includegraphics[width=3.0in]{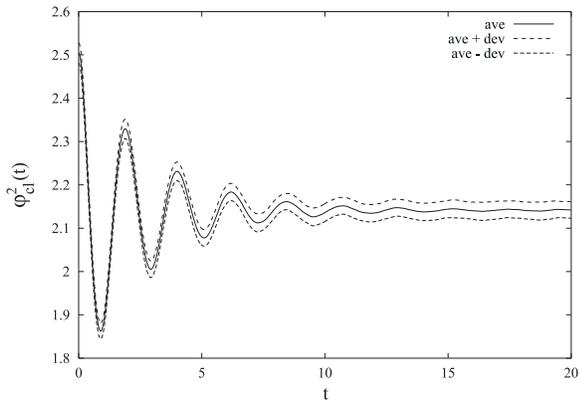}
   \caption{Monte-Carlo calculation of $\phi^2_{\text{cl}}(t)$
   versus $t$.}
   \label{f:phi}
\end{figure}
\begin{figure}[!h]
   \centering
   \includegraphics[width=3.0in]{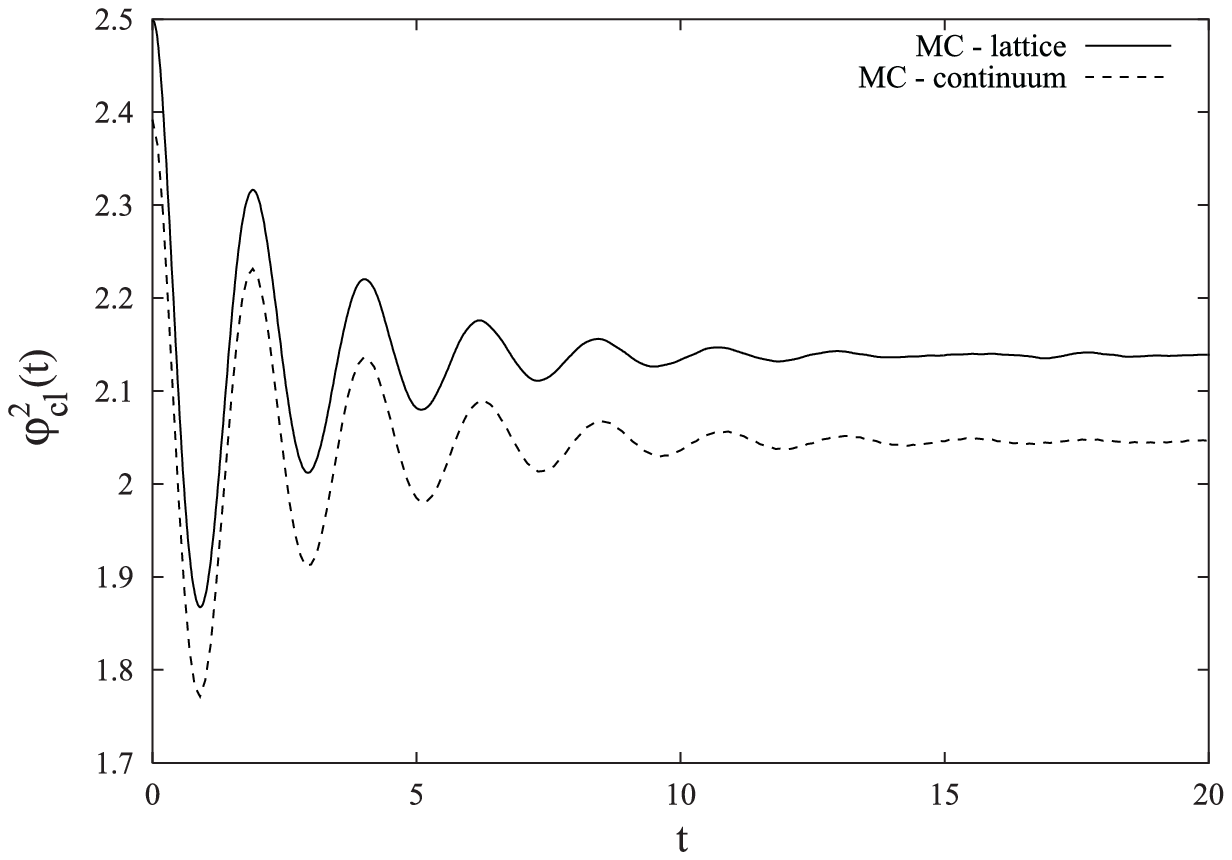}
   \caption{Comparison of the \emph{lattice} and the \emph{continuum}
            Monte Carlo results, respectively.}
\label{f:lat_cont}
\end{figure}
\begin{figure}[!h]
   \centering
   \includegraphics[width=3.0in]{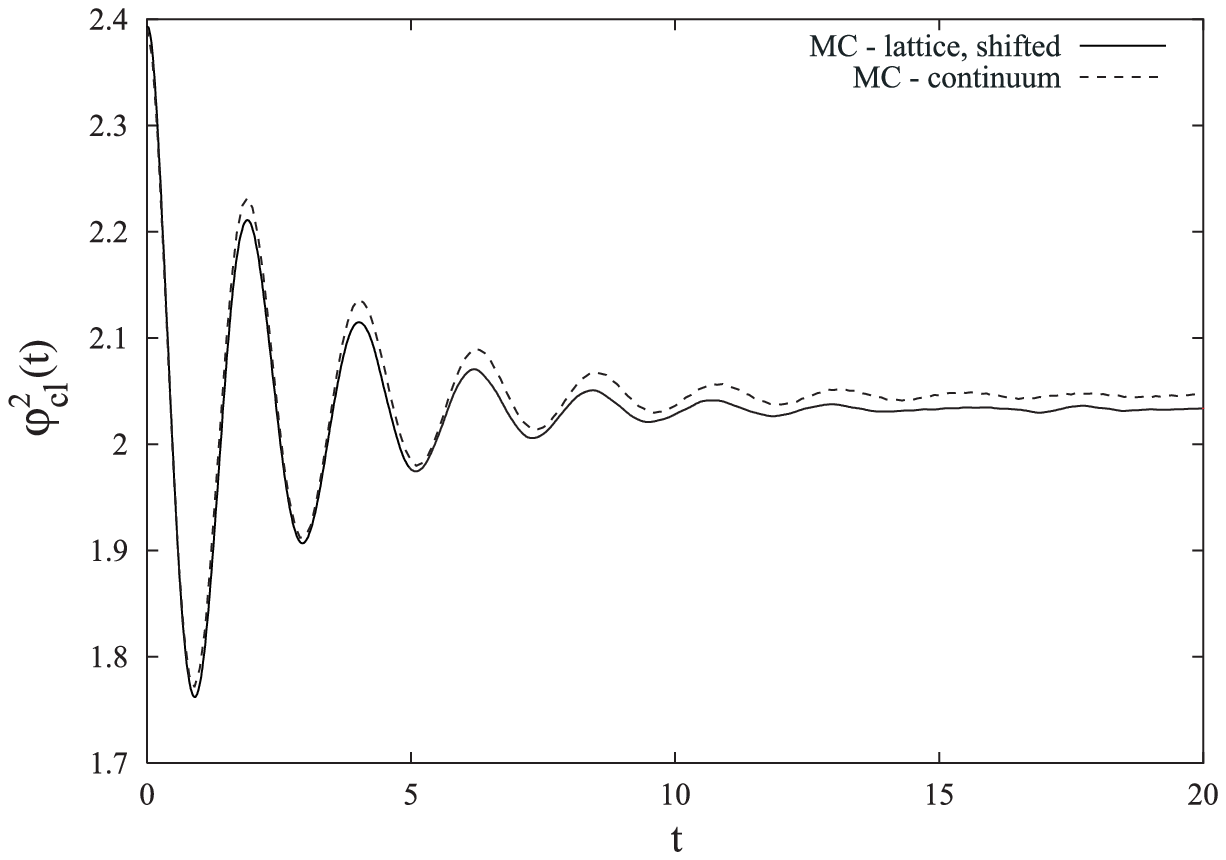}
   \caption{Comparison of the \emph{shifted lattice} and the \emph{continuum}
            Monte Carlo results, respectively.}
\label{f:lat_shift}
\end{figure}

In Fig.~\ref{f:lat_cont} we compare the calculated time dependence
of $\phi^2_{\text{cl}}(t)$, obtained using the periodic and
momentum space, respectively. In fact, the main source of
difference resides in the different initial values. To illustrate
this aspect we depict the same numerical results in
Fig.~\ref{f:lat_shift}, but shift the lattice curve such that we
match the initial value of $\phi^2_{\text{cl}}(t)$.

%
%
\section{Conclusions}

In this paper we have discussed two approaches of obtaining the
dynamical evolution of a classical system, one based on a lattice
formulation in coordinate space, the other in momentum space. Both
methods require the assumption of periodic boundary conditions,
but the different levels at which this assumption is made, allows
the momentum space approach to avoid certain artifacts of the
lattice based method. In particular the intrinsic mismatch in
initial conditions at finite cut-off values, results in different
values of the ``thermalized'' field, at large times. The
discrepancy is worse for smaller values of the cut-off, but the
two approaches converge to the same result in the continuum limit.
The mismatch in initial conditions is due to the fact that by
using a finite difference approximation for the spatial derivative
operator together with the assumption of periodic boundary
conditions on the lattice, we have in fact introduced a
approximation of the dispersion relation (see
Eq.~(\ref{e:omegak_lattice}) -- lattice, and Eq.~(\ref{e:omegak})
-- continuum), which is now viewed as an expansion in the lattice
spacing $a$. In order to improve the quality of the spatial
derivative approximation in the lattice case, one would normally
have to take the limit when the lattice spacing $a$ goes to zero.
We are however prevented from doing that, since the choice of the
momentum cut-off $\Lambda$ also determines the choice of the
lattice spacing $a=\pi/\Lambda$. Consequently we cannot improve
the agreement of the lattice dispersion relation with the
continuum for a given momentum space cut-off. The momentum  space
(continuum) approach does not exhibit this limitation.

One may think of modifying the unperturbed Hamiltonian in order to
effectively obtain a higher-order approximation of the dispersion
relation while still having the same type of equations of motion,
similar to the improved action framework in lattice QCD. This
would result in new values $\hat q^2$, and would require
cancelling the various orders of $a$ in a rigorous fashion. (The
standard lattice calculation introduces values of $\hat q^2$ which
differ from $q^2$ already at order $a^2$.) However, this is beyond
the scope of the present work, and since we are in fact able to
obtain an exact solution for the continuum problem, we merely
state here the differences between the lattice and the continuum
approach.

%
%
\begin{acknowledgments}
Present calculations are made possible by grants of time on the
parallel computers of the Mathematics and Computer Science
Division, Argonne National Laboratory. The work of BM was
supported by the U.S. Department of Energy, Nuclear Physics
Division, under contract No. W-31-109-ENG-38.
\end{acknowledgments}


\begin{thebibliography}{99}

\bibitem{ref:abw}
   G. Aarts, G.F. Bonini, and C. Wetterich,
   Phys.\ Rev.\ D {\textbf 63}, 025012 (2001)
   [hep-ph/007357].

\bibitem{ref:bva_cl}
   K.~Blagoev, J.~F.~Dawson,  F.~Cooper, and B.~Mihaila,
   ``Schwinger-Dyson approach to non-equilibrium classical field theory,''
   Phys.\ Rev.\ D (in press)
   [hep-ph/0106195].

\bibitem{ref:newberges}
   J. Berges,
   ``Controlled nonperturative dynamics of quantum fields out of equilibrium,''
   [hep-ph/0105311].

\bibitem{ref:BMIM}
   B.~Mihaila and I.~Mihaila,
   ``Numerical approximations using Chebyshev polynomial expansions,''
   [physics/9901005].

\bibitem{ref:MDC}
   B.~Mihaila, J.~F.~Dawson, and F.~Cooper,
   Phys.\ Rev.\ D {\textbf 56} (1997) 5400
   [hep-ph/9705354].

\bibitem{ref:paper1}
   B.~Mihaila, T.~Athan, F.~Cooper, J.~F.~Dawson, and S.~Habib,
   Phys.\ Rev.\ D {\textbf 62} (2000) 125015
   [hep-ph/0003105].

\bibitem{ref:us1}
   B.~Mihaila,  F.~Cooper, and  J.~F.~Dawson
   Phys.\ Rev.\ D {\textbf 63} (2001) 096003
   [hep-ph/ 0006254].

\bibitem{ref:lin_sig}
   F.~Cooper, Y.~Kluger, E.~Mottola and J.~P.~Paz,
   Phys.\ Rev.\ D {\textbf 51} (1995) 2377;
   M.~A.~Lampert, J.~F.~Dawson and F.~Cooper,
   Phys.\ Rev.\ D {\textbf 54} (1996) 2213.

\end{thebibliography}
\end{document}